\documentclass{JHEP}
\preprint{ KUL-TF-99/17 \\ DFTT 28/99 }
\title{Ramond-Ramond couplings of non-BPS D-branes\thanks{
Work supported by the European Commission TMR programme ERBFMRX-CT96-0045.}}
\author{Marco Bill\'o$^{a}$, Ben Craps\thanks{
Aspirant FWO, Belgium} \ and Frederik Roose \\
Instituut voor theoretische fysica \\
Katholieke Universiteit Leuven, B-3001 Leuven, Belgium \\

$^a$ Dipartimento di Fisica Teorica, Universit\`a di
Torino and \\
I.N.F.N., Sezione di Torino, via P. Giuria 1, I-10125, Torino, Italy \\
E-mail:
\email{billo@to.infn.it, Ben.Craps@fys.kuleuven.ac.be,}\\
\email{Frederik.Roose@fys.kuleuven.ac.be}}
\abstract{
We study how non-BPS type II D-branes couple to R-R potentials. Upon tachyon condensation the couplings
we find give rise to the Wess-Zumino action of BPS D-branes.}

\newcommand{\eq}[1]{Eq.~(\ref{#1})}

\def\beq{\begin{equation}}
\def\eeq{\end{equation}}
\def\beqa{\begin{eqnarray}}
\def\eeqa{\end{eqnarray}}

\newcommand{\EQ}{\begin{equation}}
\newcommand{\EN}{\end{equation}}
\newcommand{\bea}{\begin{eqnarray}}
\newcommand{\ena}{\end{eqnarray}}

\renewcommand{\a}{\alpha}

\renewcommand{\thefootnote}{\fnsymbol{footnote}}
\def\one{{\hbox{ 1\kern-.8mm l}}}

\def\ii{{\rm i}}

\newlength{\bredde}
\def\slash#1{\settowidth{\bredde}{$#1$}\ifmmode\,\raisebox{.15ex}{/}
\hspace*{-\bredde} #1\else$\,\raisebox{.15ex}{/}\hspace*{-\bredde} #1$\fi}

\newsavebox{\uuunit}
\sbox{\uuunit}
    {\setlength{\unitlength}{0.825em}
     \begin{picture}(0.6,0.7)
        \thinlines
        \put(0,0){\line(1,0){0.5}}
        \put(0.15,0){\line(0,1){0.7}}
        \put(0.35,0){\line(0,1){0.8}}
       \multiput(0.3,0.8)(-0.04,-0.02){12}{\rule{0.5pt}{0.5pt}}
     \end {picture}}

\newsavebox{\zzzbar}
\sbox{\zzzbar}
  {\setlength{\unitlength}{0.9em}
  \begin{picture}(0.6,0.7)
  \thinlines
  \put(0,0){\line(1,0){0.6}}
  \put(0,0.75){\line(1,0){0.575}}
  \multiput(0,0)(0.0125,0.025){30}{\rule{0.3pt}{0.3pt}}
  \multiput(0.2,0)(0.0125,0.025){30}{\rule{0.3pt}{0.3pt}}
  \put(0,0.75){\line(0,-1){0.15}}
  \put(0.015,0.75){\line(0,-1){0.1}}
  \put(0.03,0.75){\line(0,-1){0.075}}
  \put(0.045,0.75){\line(0,-1){0.05}}
  \put(0.05,0.75){\line(0,-1){0.025}}
  \put(0.6,0){\line(0,1){0.15}}
  \put(0.585,0){\line(0,1){0.1}}
  \put(0.57,0){\line(0,1){0.075}}
  \put(0.555,0){\line(0,1){0.05}}
  \put(0.55,0){\line(0,1){0.025}}
  \end{picture}}

\begin{document}

\renewcommand{\thefootnote}{\arabic{footnote}}
\setcounter{footnote}{0}
In a seminal series of papers, Sen has shown how to obtain an exact conformal 
field theory description of non-BPS D-branes (see \cite{sen} for two of the 
main papers and two recent reviews). 
In this setup, a non-BPS D($p-1$)-brane remains after
tachyon condensation on a D$p$--anti-D$p$-brane pair. 
In type II theories the resulting object still
has a tachyonic mode on its worldvolume, but performing the orientifold to 
type I stabilizes the non-BPS brane. 
In a related development, D-brane charges have been shown to take values in
appropriate K-theory groups of space-time. A major result is that all
lower-dimensional D-branes can be considered in a unifying manner as 
non-trivial excitations on the appropriate configuration of higher-dimensional 
branes. For type IIB this was demonstrated by Witten in \cite{witten}, where 
all branes are built from sufficiently many D9--anti-D9
pairs. As to type IIA, Horava outlined the construction of lower branes from 
non-BPS D9-branes \cite{horava}. 

As is well-known, type II BPS D-branes couple to Ramond-Ramond gauge fields 
through the Wess-Zumino action \cite{GHM} (see \cite{anomalous} for
string computations checking the form of this action):
\beq \label{wzbps}
S_{\rm WZ}=\frac{T_p}{\kappa}\int_{p+1} C\wedge {\rm Tr}\, {\rm e}^{2\pi\a '\,F+B}\wedge
\sqrt{\hat{A}(R_T)/\hat{A}(R_N)}~.
\eeq
Here $T_p/\kappa$ denotes the D$p$-brane tension, $C$ a formal sum of R-R 
potentials, $F$ the gauge field on  the brane and $B$ the NS-NS two-form.
The trace is over the Chan-Paton indices.
Further, $R_T$ and $R_N$ are the curvatures of the tangent and normal 
bundles of the D-brane world-volume, and $\hat{A}$ denotes the A-roof genus.
In the setup of Ref.~\cite{witten}, where one starts with an unstable configuration of 
supersymmetric branes 
and anti-branes, the coupling on the BPS-brane that remains after tachyon condensation 
is inherited from the similar coupling of the parent branes. 
In the scenario of Ref.~\cite{horava}, one starts from non-BPS branes, which
can decay into a lower BPS brane via tachyon condensation.
In this case, it has  not been clear how the resulting objects acquire 
the desired couplings in \eq{wzbps}. 
In this note we argue that all type II non-BPS branes couple 
universally to Ramond-Ramond fields as given by
\beq \label{wznonbps}
S'_{\rm WZ}=a\int_{p+1} C\wedge d\,{\rm Tr}\, T \, {\rm e}^{2\pi\a '\,F+B}\wedge
\sqrt{\hat{A}(R_T)/\hat{A}(R_N)}~,
\eeq
where $T$ is the real, adjoint tachyon field living on the non-BPS brane and 
$a$ is a constant.\footnote{This constant will be fixed in the next paragraph by imposing that the
BPS D($p-1$)-brane we find there have the expected R-R charge. Then this action predicts the R-R charges
of the lower BPS D-branes that can be constructed from the non-BPS D$p$-brane.}
One term of this action (the one describing the coupling of a non-BPS D$p$-brane 
to $C_p$) was discussed in Refs.~\cite{sen, horava}. 
Below we will show how, upon tachyon condensation,
these non-BPS ``Wess-Zumino'' couplings induce the appropriate Wess-Zumino
action for the resulting BPS-branes. The cases D$p$ $\rightarrow$ D($p-1$) 
and D$p$ $\rightarrow$ D($p-3$) will be treated in detail. It will turn out, for instance,
that the \mbox{R-R} charges of the D8-branes and D6-branes one constructs from unstable D9-branes 
\cite{horava} have the expected ratio.
Moreover, we check the presence of these R-R couplings by performing 
various disc amplitudes with an open string tachyon inserted at the boundary.
\paragraph{Relation to Wess-Zumino action} In Ref.~\cite{horava} 
Horava described how to construct BPS D($p-2k-1$)-branes as 
bound states of (sufficiently many) unstable D$p$-branes. The lower-dimensional BPS
branes arise as the result of the condensation of a tachyon field 
into a vortex configuration, accompanied by non-trivial gauge fields.
We indicate now how the R-R couplings that we propose in \eq{wznonbps} account 
for the R-R couplings (\ref{wzbps}) that the stable lower-dimensional brane 
emerging from the condensation must possess.
\par
Consider first a single non-BPS D$p$-brane. There is a  real
tachyon field living on its worldvolume. The tachyon potential 
is assumed to be such that the vacuum manifold consists of the two points 
$\{T_0,-T_0\}$.\footnote{The symbol $T_0$ should not be confused with the tension of a D0-brane, which
will never explicitly appear in this paper.}
The tachyon 
can condense to a non-trivial (anti)-kink configuration $T(x)$ depending on a
single coordinate. 
The R-R coupling (\ref{wznonbps}) on the D$p$-brane reads in this case 
\begin{equation}
\label{m1}
a\int_{p+1} C\wedge dT\wedge {\rm e}^{2\pi\alpha' F + B} \wedge 
\sqrt{\hat{A}(R_T)/\hat{A}(R_N)}~,
\end{equation}
the first term of which was suggested in Ref. \cite{horava} and shown to be present by a 
disc computation (in an alternative formalism) in Ref.~\cite{sen}.
It involves the topological density  $\partial_x T(x)$, which is localized at 
the core of the kink and is such that $\int dT(x) = \pm 2 T_0$. 
In the limit of zero size we would have $dT(x) = 2 T_0\delta(x-x_0) dx$, and the above 
action would take the form\footnote{%
Actually, trying to follow the reduction of a D$p$-brane
to a lower-dimensional one, there is a puzzle  
concerning the gravitational part
$\sqrt{\hat{A}(R_T)/\hat{A}(R_N)}$. 
The directions along the parent brane transverse to the 
smaller brane contribute originally to $\hat A(R_T)$. It is not clear to us how they are 
reassigned to the normal bundle in the reduced action.
In fact, this problem seems also to be present for the reduction of brane-antibrane pairs
to lower-dimensional BPS branes as in Ref.~\cite{witten}, where only the standard 
WZ actions (\ref{wzbps}) are involved.} 
of the usual Wess-Zumino
effective action for a BPS D($p-1$)-brane, localized in the $x$-direction at
$x_0$:
\begin{equation}
\label{m2}
2T_0a\int_{p} C\wedge {\rm e}^{2\pi\alpha' F + B} 
\wedge \sqrt{\hat{A}(R_T)/\hat{A}(R_N)}~.
\end{equation}
In reality, the D($p-1$)-brane will have a certain thickness in the direction of the kink.
\par
Note that the constant $a$ can be fixed in terms of $T_0$ by equating $2T_0a$ with the tension 
$T_{p-1}/\kappa$ of a BPS D($p-1$)-brane. This being done, the remainder of this paragraph provides a
non-trivial check on our couplings in \eq{wznonbps}.  
\par
As a less trivial example, 
let us start from two unstable D$p$-branes. The tachyon field $T$,
transforming in the adjoint of the ${\rm U}(2)$ gauge group, 
can form a non-trivial vortex configuration in co-dimension three.
The tachyon potential is assumed to be such that the minima of $T$ have 
the eigenvalues $(T_0,-T_0)$, so that the vacuum manifold is 
${\cal V} = {\rm U}(2)/({\rm U}(1)\times {\rm U}(1))=S^2$. The possible
stable vortex configurations  $T({\bf x})$, depending on $3$ coordinates 
$x^i$ transverse to the ($p-2$)-dimensional core of the vortex, are classified by the
non-trivial embeddings of the ``sphere at infinity'' $S^{2}_\infty$ into the
vacuum manifold, namely by $\pi_{2}({\cal V}) = {\bf Z}$.

Apart from  the ``center of mass'' ${\rm U}(1)$ 
subgroup we are in the situation of the Georgi-Glashow model,
where the tachyon field $T({\bf x})= T^a({\bf x}) \sigma^a$ 
($\sigma^a$ being the Pauli matrices) sits in the adjoint of  ${\rm SU}(2)$, 
and the vacuum manifold is described by $T^a T^a = T_0^2$.
The vortex configuration of winding number one, which is the 't Hooft-Polyakov
monopole, is of the form
\begin{equation}
\label{m3}
T({\bf x}) = f(r) \sigma_a x^a~,
\end{equation}
where $r$ is the radial distance in the three transverse directions, 
and the prefactor $f(r)$ goes to a constant for $r\to 0$ and approaches
$T_0/r$ for $r\to\infty$.

The finite energy requirement implies that $D_i T^a$ 
vanishes sufficiently fast at infinity, from which it follows that a vortex is accompanied by a 
non-trivial gauge field; for the case (\ref{m3}) above, the non-trivial 
part of the SU(2) gauge field has the form
\begin{equation}
\label{m4}
{\cal A}^a_i({\bf x}) = h(r) \epsilon^a{}_{ij} x^j~,
\end{equation} 
with $h(r)$ approaching a constant for $r\to 0$, while $h(r)\sim 1/r^2$ 
at infinity.
The field-strength in the unbroken ${\rm U}(1)$
direction,
\begin{equation}
\label{m5}
{\cal G}_{ij} = {T^a\over T_0} {\cal F}^a_{ij}~~,
\end{equation}
corresponds to a
non-trivial U(1) bundle on the sphere at infinity, {\it i.e.} the magnetic 
charge $g = \int_{S^{2}_\infty} {\cal G}$ is non-zero (and in fact equals 
the winding number of the vortex in appropriate units).
Thus there is a magnetic charge density 
in the transverse directions, defined by $d{\cal G} = \rho({\bf x}) d^3x$, 
which is concentrated at the core of the vortex 
solution. In the zero size limit,  there would be a point-like magnetic charge
at the location of the core: $\rho({\bf x}) = g\, \delta^3({\bf x} - {\bf x}^0)$.
\par
The  WZ action (\ref{wznonbps}) for the D$p$-brane can be rewritten as
\begin{equation}
\label{m6}
a\int_{p+1} C\wedge d\,{\rm Tr}\{ T\, {\rm e}^{2\pi\alpha' {\cal F}}\}
\wedge{\rm e}^{2\pi\alpha' {\hat F} + B} \wedge 
\sqrt{\hat{A}(R_T)/\hat{A}(R_N)}~,
\end{equation}
where we have split the U(2) field-strength into its SU(2) part ${\cal F}$
and its U(1) part $\hat F$. Inserting the 't Hooft-Polyakov configuration
for the tachyon and the SU(2) gauge field, we see that Eq.~(\ref{m5}) involves 
precisely the magnetic monopole field ${\cal G}= T^a {\cal F}^a$; we get indeed
\begin{eqnarray}
\label{m7}
& & 2\pi\alpha' a\int_{p+1}  C\wedge d{\rm Tr}\{ T\,{\cal F}\}\wedge
{\rm e}^{2\pi\alpha' {\hat F} + B} \wedge 
\sqrt{\hat{A}(R_T)/\hat{A}(R_N)}\nonumber\\
&=& 2\pi\alpha' a T_0
\int_{p+1}  C\wedge \rho({\bf x}) d^3x\wedge
{\rm e}^{2\pi\alpha' {\hat F} + B} \wedge 
\sqrt{\hat{A}(R_T)/\hat{A}(R_N)}~.
\end{eqnarray}
Thus we have a distribution of D($p-3$)-brane charge localized at the 
core of the vortex; in particular, in the limit of zero-size core we recover
the R-R couplings (\ref{wznonbps}) of a BPS D($p-3$)-brane that supports the 
U(1) gauge field $\hat F$.
\par
Since the minimal magnetic charge $g$ is $4\pi$ in our units, Eq.~(\ref{m7})
and the remark after \eq{m2} lead to the expected ratio $4\pi^2\alpha'$ for the R-R charges of 
D($p-3$)-~and D($p-1$)-branes.
\par
The mechanism described above generalizes
to the reduction of a non-BPS D$p$-brane to a D$(p-2k-1)$-brane via
tachyon condensation, described in \cite{horava}. In this case, it is 
convenient to start with $2^k$ unstable D$p$-branes. The configuration of 
vorticity one for the tachyon field, which sits in the adjoint of U$(2^k)$, is 
of the form 
\begin{equation}
\label{m8}
T({\bf x}) = f(r) \, \Gamma_i x^i~,
\end{equation}  
where $r$ is the radius in the $2k+1$ transverse dimensions $x^i$, and the
$\Gamma$-matrices in these dimensions are viewed as 
$U(2^k)$ elements. Eq.~(\ref{m8}) is a direct generalization of the 't Hooft-%
Polyakov case, Eq.~(\ref{m3}). Again, the finite energy requirement should imply
a non-trivial gauge field configuration, leading to a non-zero 
generalized magnetic charge $\int_{S^{2k}_\infty}{\rm Tr}\{T ({\cal F})^k\}$.
In such a background, the WZ action (\ref{wznonbps}) contains the factor
$d\,{\rm Tr}\{T {\cal F}^k\}= \rho({\bf x}) d^{2k+1}x$; the (generalized) magnetic
charge density $\rho$ is concentrated at the core of the vortex, and in the
zero-size limit reduces to a delta-function in the transverse space. Thus we
are left with the WZ action for a D$(p-2k-1)$-brane.    
\paragraph{String computation}
To compute the disc scattering amplitudes necessary to check \eq{wznonbps}, it is convenient to
conformally map the disc to the upper half plane and use the ``doubling trick'' as described, for
instance, in Ref.~\cite{hashimoto}. This trick consists in replacing, {\it e.g.}, $\bar X^{\mu}
(\bar z)$ by $S^\mu_{~\nu} X^\nu(\bar z)$, where $S^{\mu}_{~\nu}$ is diagonal, with entries $1$ in the 
worldvolume and $-1$ in the transverse directions, and then treating the fields depending on 
$\bar z$  as if $\bar z$ were a holomorphic variable living on the lower half plane. The fermionic 
$\psi^\mu$ fields are treated in
the same way. As to the spin fields in the \mbox{R-R} sector, for BPS D$p$-branes in type IIA 
$\bar S^{\dot\alpha}(\bar z)$ is replaced by 
$(\gamma^0\gamma^1\cdots\gamma^p)^{\dot\alpha}{}_\beta\,S^\beta(\bar z)$ (where the chirality flips
because $p$ is even). For type IIB  $\bar S^{\alpha}(\bar z)$ is replaced by
$(\gamma^0\gamma^1\cdots\gamma^p)^{\alpha}{}_\beta\,S^\beta(\bar z)$, where now $p$ is odd. For the
non-BPS D$p$-branes we are studying here, $p$ is odd in IIA and even in IIB, so that there is a
chirality flip in IIB and not in IIA. The explicit computations below will be done for IIA, but the
story is, of course, completely analogous for IIB.

The first amplitude we are going to compute is the two point function of one open string tachyon and
a R-R potential in the presence of a single non-BPS D$p$-brane in IIA.\footnote{This has been done before in a
formalism in which non-BPS D-branes are constructed in an alternative way \cite{sen}.}
This will establish the first term in the 
expansion
of \eq{wznonbps}. We take the R-R vertex operator in the ($-1/2,-1/2$) picture (which
exhibits the R-R field strengths rather than the potentials):
\beq
V_{RR}=H_{\alpha\dot\beta}\, S^\alpha(z)\, \bar S^{\dot\beta}(\bar z)\,e^{\ii k\cdot X(z,\bar z)}
\rightarrow H_{\alpha\dot\beta}\, S^\alpha(z)\, (\gamma^0\gamma^1\cdots\gamma^p)
^{\dot\beta}{}_{\dot\gamma}\,S^{\dot\gamma}(\bar z)\,e^{\ii k\cdot X(z)}\,
e^{\ii k\cdot S\cdot X(\bar z)}~,
\eeq
where $H_{\alpha\dot\beta}$
is the bispinor containing the R-R field strengths
and $k$ the momentum of the R-R
potential. We have omitted the superghost part and do not keep track of the overall normalization,
since we are not able to directly determine the constant $a$ in \eq{wznonbps} anyway.\footnote%
{However, we will be
interested in the relative normalization of this amplitude with respect to the ones with photons
inserted. The constant $a$ itself was fixed in the previous paragraph.}
The tachyon vertex operator is put in the $-1$ picture:
\beq
V_T=T(k')\,e^{\ii k'\cdot X(y)}~~,
\eeq 
where $T$ and $k'$ are the tachyon polarization and momentum and $y$ is a point on the real axis. 
Again, the superghost part
is not displayed. The three insertion points $z$, $\bar z$ and $y$ can be fixed by introducing ghost
fields. Then the contributions of the ghost, superghost and $X$ sectors combine into $(z-\bar
z)^{5/4}$. The contraction of the two spin fields in the fermionic sector gives
\beq
<S^\alpha(z)\,S^{\dot\gamma}(\bar z)>=(z-\bar z)^{-5/4}\,C^{\alpha\dot\gamma}~~,
\eeq
with $C$ the charge conjugation matrix. 
The amplitude becomes 
\beq
T\,H_{\alpha\dot\beta}
(\gamma^0\gamma^1\cdots\gamma^p)^{\dot\beta}{}_{\dot\gamma}\,C^{\alpha\dot\gamma}\times K~~,
\eeq
where $K$ is a global factor.
Tracing over the spinor indices, only the part of $H_{\alpha\dot\beta}$
proportional to $H_{\mu_1\ldots\mu_{p+1}} (C\gamma^{\mu_1\ldots\mu_{p+1}})_{\alpha\dot\beta}$
contributes, making the amplitude proportional to
$T\, H_{\mu_1\ldots\mu_{p+1}}\epsilon^{\mu_1\ldots\mu_{p+1}}$. Upon integration by parts,
this confirms the first term of~\eq{wznonbps}.

There is a kinematical subtlety in this computation. String scattering amplitudes can only be
computed for on-shell external particles. It is easy to convince oneself that, since the tachyon
carries only momentum along the brane and the momentum along the brane is conserved, the tachyon and
the R-R potential cannot be both on-shell. As a way out, one could consider branes with Euclidean
signature, for which this kinematical problem does not occur, and then extrapolate the couplings one
finds there to their Minkowski cousins.

To check the second term of \eq{wznonbps}, depending linearly on $F$, we add to the previous
amplitude a vertex operator for a gauge field. This vertex operator is in the 0 picture:
\beq\label{photon}
V_A=A_\mu(\ii\dot X^\mu(w)+2\alpha'\,p\cdot\psi\psi^\mu(w))\,e^{\ii p\cdot X(w)}~~,
\eeq
where this time we have kept track of all normalization factors. Here $A_\mu$ is the polarization of
the gauge field, $p$ is its momentum and $w$ is on the real axis. Only the fermionic part of the 
photon vertex operator
can lead to terms of the type we are looking for (the photon should provide two gamma-matrices), so
we ignore possible contributions from the bosonic part. We will compute the amplitude to lowest order 
in the photon momentum. This means that we will put $p$ equal to zero in the bosonic sector,
thus keeping only the explicit $p$ dotted with a $\psi$ in \eq{photon}. We follow the previous
computation as closely as possible by fixing again $z$, $\bar z$ and $y$, such that only $w$ needs
to be integrated over. In the limit of small photon momentum the ghost, superghost and $X$ sector
contributions are unchanged (they multiply to $(z-\bar z)^{5/4}$). The fermionic correlator is
\beq
2\alpha' p_\nu A_\mu <S^\alpha(z)\,\psi^\nu\psi^\mu(w)\,S^{\dot\gamma}(\bar z)>=
-\ii\,\alpha' p_\nu A_\mu\,(\gamma^{\nu\mu})^{\alpha\dot\gamma}\, (w-z)^{-1}(w-\bar z)^{-1}
(z-\bar z)^{-1/4}.
\eeq
The resulting integral can be done by a contour integration:
\beq
(z-\bar z)\int_{-\infty}^{+\infty}\,dw\,(w-z)^{-1}(w-\bar z)^{-1}=2\pi\ii~~,
\eeq
leading to
\beq
2\pi\alpha'\, p_\nu\,A_\mu\,T\,H_{\alpha\dot\beta}\,
(\gamma^{\nu\mu})^{\alpha\dot\gamma}(\gamma^0\gamma^1\cdots\gamma^p)^{\dot\beta}{}_{\dot\gamma}\,
\times K
\eeq
for the amplitude. This corresponds indeed to the term in \eq{wznonbps} linear in $F$. Note that the
factor $2\pi\alpha'$ multiplying $F$ in \eq{wznonbps} comes out correctly.

The generalization to multiple (low momentum) photon insertions is straightforward. The dependence 
on the photon insertion points of the relevant part of the fermionic correlator factorizes, such
that each integration reduces to the one-dimensional integral described in the previous paragraph.
It is also easy to include Chan-Paton factors in the above computations, leading to the trace in
\eq{wznonbps}. Finally, one could check the presence of the gravitational terms in \eq{wznonbps}
explicitly. Since all graviton vertex operators can be inserted in the ($0,0$) picture, the various
contractions will be identical to the ones used in Ref.~\cite{anomalous}. 

Note that, from a technical point of view, the only role of the tachyon in the above computations 
is to provide its superghost part, allowing one to insert the R-R vertex operator in the $(-1/2,-1/2)$
picture, instead of the ($-3/2,-1/2$) one. Thus the inclusion of the tachyon proves wrong one's first 
impression that non-BPS D-branes cannot couple to the closed string R-R sector because of the 
GSO-projection. Apart from this, the above computations perfectly parallel their counterparts for
BPS D-branes.     

\paragraph{Acknowledgement}
We would like to thank Pieter-Jan De Smet and Walter Troost for discussions.

\end{document}